\documentstyle[apjfonts, emulateapj]{article}
\submitted{To appear in the Astrophysical Journal, March 20, 2001 issue}
\lefthead{KIM, ZUCKERMAN, \& SILVERSTONE}
\righthead{GIANTS WITH EXCESS FAR--IR EMISSION}

\def\spose#1{\hbox to 0pt{#1\hss}}
\newcommand\lsim{\mathrel{\spose{\lower 3.0pt\hbox{$\mathchar"218$}}
     \raise 2.0pt\hbox{$\mathchar"13C$}}}
\newcommand\gsim{\mathrel{\spose{\lower 3.0pt\hbox{$\mathchar"218$}}
     \raise 2.0pt\hbox{$\mathchar"13E$}}}

\begin{document}
\title{EXTENT OF EXCESS FAR INFRARED EMISSION AROUND LUMINOSITY CLASS III STARS}
\author{Sungsoo S. Kim, B. Zuckerman, \& Murray Silverstone\altaffilmark{1}}
\affil{Division of Astronomy \& Astrophysics, University of California,
Los Angeles, CA 90095-1562;\\
sskim@astro.ucla.edu, ben@astro.ucla.edu, murray@as.arizona.edu}

\altaffiltext{1}{Present address: Steward Observatory, University of Arizona,
Tucson, AZ 85721}

\begin{abstract}
With the {\it Infrared Space Observatory}, we conducted $3\times 3$-pixel
imaging photometry of twelve luminosity class III stars, which were
previously presumed to have dust particles around them, at far infrared
wavelengths (60 and $90 \, {\rm \mu m}$).  Eleven out of twelve targets
show a peak of excess (above photosphere) far infrared emission
at the location of the star, implying that the dust particles are truly
associated with stars.  To estimate the size of the excess emission source,
the flux ratio of center to boundary pixels of the $3\times 3$ array was
examined.
The radius of the dust emission
is found to be $\sim 3000$ to $\sim 10000$~AU for a thin shell
distribution, and $\sim 5000$ to $\sim 25000$~AU for a uniform distribution.
We consider three models for the origin of the dust:
disintegration of comets, sporadic dust
ejection from the star, and emission from nearby interstellar cirrus.
The data seem to rule out the first model (as far as the ``Kuiper--belt'' like
particles are assumed to be large blackbody grains), but do not enable
us to choose between the other two models.
\end{abstract}


\section{INTRODUCTION}
\label{sec:introduction}

\markcite{ZKL95}Zuckerman, Kim, \& Liu (1995) correlated
the Bright Star Catalog (\markcite{HW91}Hoffleit \& Warren 1991)
and the Michigan Spectral Catalog (\markcite{HCS75}Houk, Cowley, \&
Smith-Moore 1975--1988) with the IRAS
catalogs to determine which, if any, luminosity class III giant stars
(first ascent red giants) have associated circumstellar dust particles
that radiate at far infrared (far--IR) wavelengths.  Of more than
40,000 class III giant stars in the two catalogs, they found that
perhaps $~300$ have associated dust.

Whereas the presence of particulate material near pre--main sequence,
main--sequence, AGB, and supergiant stars may be accounted for in rather
straightforward ways (remnants from the star formation for the former
two, and mass loss for the latter two), there is no obvious way to
produce and retain large amounts of dust near first ascent giant stars,
which are thought to be too old to still possess much left--over material and
insufficiently evolved to have lost a significant amount of their mass.
However, one may
conjecture that the presence of dust around first ascent giants may
involve one or more of the following phenomena: mass loss, binarity,
planetary systems, and evaporation of Kuiper--belt material.
Recently, \markcite{J99}Jura (1999) has analysed
three models for the dust around eight first ascent giant stars:
orbiting dust resulting from the disintegration of comets located in
extrasolar analogs of the Kuiper belt,
dust sporadically ejected from the star, and emission
from particles in the interstellar medium which are accidentally
near the star, the ``cirrus hotspot''.

Among the \markcite{ZKL95}Zuckerman et al. (1995) list of 300,
we observed 12 first ascent giant stars
with the PHOT instrument of the {\it Infrared Space Observatory}
({\it ISO}\footnote{{\it ISO} is an ESA project with instruments funded by
ESA member states (especially the PI countries: France, Germany,
The Netherlands and the United Kingdom) and with the participation
of ISAS and NASA}) to determine
if the excess far--IR emission is truly associated with the stars and to
estimate the size of the region that produces the excess far--IR emission.

\S~\ref{sec:observation} describes our {\it ISO} observations and data
reduction.  \S~\ref{sec:analysis} discusses the data analysis including
the determination of the size of the far--IR excess source.
\S~\ref{sec:discussion} then compares our results with Jura's models.
A summary is given in \S~\ref{sec:summary}.

\section{OBSERVATIONS \& DATA REDUCTION}
\label{sec:observation}

The target stars were observed with the C100 camera of the ISOPHOT
instrument using filters 60 ($\lambda_c =60.8 \, {\rm \mu m}$,
$\Delta \lambda = 23.9 \, {\rm \mu m}$) and 90 ($\lambda_c =95.1 \,
{\rm \mu m}$, $\Delta \lambda = 51.4 \, {\rm \mu m}$).  C100 is a
$3 \times 3$ Ge:Ga pixel array.  The sky coverage of each pixel is
$43.5\arcsec \times 43.5\arcsec$ and the gap between adjacent pixels
corresponds to $2.5\arcsec$ on the sky.  The diffraction limit of the ISO
is $25\arcsec$ at $60 \, {\rm \mu m}$ and $38\arcsec$ at $90 \, {\rm \mu m}$.
Observation mode PHT37--39 was used to obtain a sequence of
{\sf dark--off1--on--off2} measurements for each target and filter.
The target was measured at the {\sf on} sequence position and,
to obtain the background intensity around the target,
two off--position measurements ({\sf off1} \& {\sf off2}) were made at a
single position $2.5\arcmin$ closer to the nearer celestial pole than the
target.  Calibration measurements were performed at
the {\sf dark} and {\sf off2} sequence positions
using the Fine Calibration Source 1 (FCS1) onboard the satellite.
The integration time was 64~s for scientific measurements
(target and background) and 32~s for FCS1 measurements.

The data were reduced using the PHOT Interactive Analysis software
package (\markcite{Ge97}PIA, Gabriel et al. 1997) version 7.2.2(e).
Each measurement consisted of 64 ramps (16 for FCS1), and each ramp has
31 non--destructive readouts (63 for FCS1) and 1 destructive readout.
Data reductions described below used the default values
of the PIA except when specified.
Non--linearity correction and two--threshold deglitching were applied
to each ramp, and a linear--fit was made to the ramp to derive the
signals in V/s.  These signals were deglitched again using the minmax
clipping algorythm with a threshold set at $2.4 \sigma$.  Then reset interval
correction and dark current subtraction were applied to the signals.

\begin{figure*}
\centerline{\epsfxsize=8.8cm\epsfbox{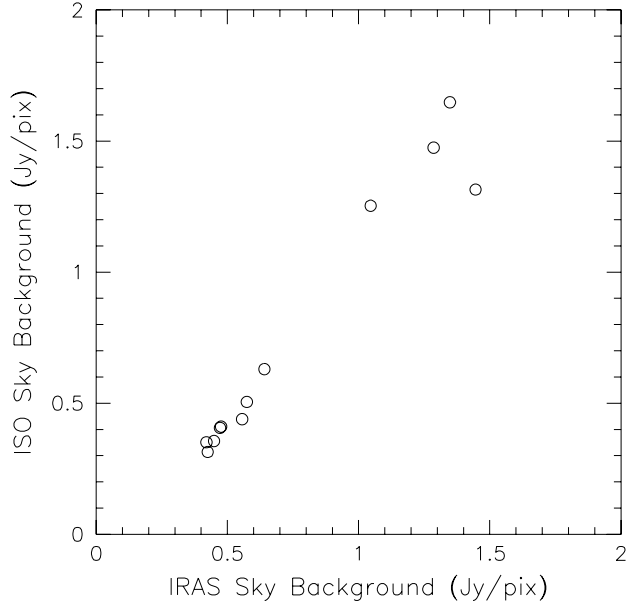}}
\caption
{\label{fig:sky}Comparison of $60 \, \mu$m background flux observed by
the {\it IRAS} and {\it ISO} near our target stars.  Fluxes are in units
of Jy per one ISO C100 pixel area.  IRAS flux is the $4\arcmin$ IRAS Sky
Survey Atlas background before subtraction of a zodiacal model.
The mean of the difference between two data sets is 0.4~\%, and
the standard deviation of the mean of the difference is 5~\%.
The uncertainty in flux ranges from 2.2 to 4.8~\% for the ISO
data, and from 0.1 to 4.6~\% for the IRAS data (the IRAS uncertainty is
from the cirrus confusion noise).
}
\end{figure*}


The responsivity of the system, that is the conversion of signal (V/s) to flux
(Jy), was obtained by time--interpolating two FCS1 measurements for each target.
At each star the background at each of the nine pixels was obtained
from an average of the {\sf off1} and {\sf off2} images.  Inspection
of the 60~$\mu$m background fluxes obtained in this way at the 12
target stars indicated a larger flux (by ~10\%) at the central pixel of the
background; this difference between central and boundary pixels is larger
than the spatial variation of the background.  We interpret the difference
as due to a bias in the FCS1 signal and correct for it as follows.  The
flux of each {\sf on} pixel is divided by the background flux of the
corresponding pixel and multiplied by the average of the 9 pixel background
flux.  This background average is then subtracted from the {\sf on} pixel
flux to produce a measure of the source flux.

To check the accuracy of data measurement and reduction, our ISO
background average and the $4\arcmin$ IRAS Sky Survey Atlas background 
(before the subtraction of a zodiacal model) were compared.
Figure~\ref{fig:sky} shows that the two flux values agree very well.

The uncertainty for the background flux was estimated by taking the rms
value of the 9 differences between 2 {\sf off} measurements of each pixel.
The uncertainty for the total boundary--pixel flux of the source was
estimated to be the square--root of 4 times the sum of the variance of
4 crossing pixels and that of 4 diagonal pixels considering the
point--symmetric nature of the Point Spread Function (PSF).  Since only
1 pixel measurement is available for the central pixel of the source,
we simply assume that background and central {\sf on}--pixel uncertainties
are proportional to their fluxes.  The average signal--to--noise ratio (S/N)
is found to be $\sim 20$ for background flux and $\sim 5$ for central
flux of the source, and $\sim 2.2$ for boundary--pixel flux of the source.
However, our analysis involves the boundary--pixel flux of the source
only in the context of a sum of 8 boundary--pixel fluxes, which has an average
S/N of $\sim 6$.

\section{DATA ANALYSIS}
\label{sec:analysis}

Measured ISO fluxes are presented in Table~\ref{table:flux} along with
the IRAS fluxes.  Since the fluxes are neither color--corrected,
nor compensated for their detector sizes and PSF shapes, the ISO
and IRAS data are not to be compared directly.

\subsection{Association with Optical Counterparts}

Since the presumed optical counterparts are targeted at the central
pixel of the $3 \times 3$ array, if the material responsible
for excess far--IR emission is distributed around the star with an intensity
concentration toward the center or with an extent smaller than one pixel
size, the intensity of the central pixel in our ISO data should be highest.
We find that the peak intensity is located at the central pixel for all
stars except HD 221776 which has $90 \, {\rm \mu m}$ peak intensity
at the boundary pixel to the south, and HD 24124 that has 60 and
$90 \, {\rm \mu m}$ peak intensities at the boundary pixel to the northwest.
We have identified in the Digitized Sky Survey a galaxy--like object at the
location of the 60 and $90 \, {\rm \mu m}$ peak intensities of HD 24124.
Since the peak intensity pixels at both passbands coincide with the
nearby galaxy, we attribute to it
the far--IR emission previously presumed to be associated with HD 24124.
However, in the case of HD 221776, because no object was found near the
position of $90 \, {\rm \mu m}$ peak intensity from the
Digitized Sky Survey, and because the $60 \, {\rm \mu m}$ peak
intensity occurs at the central pixel, we believe that the
$60 \, {\rm \mu m}$ peak intensity is truly associated with the star and
the $90 \, {\rm \mu m}$ peak intensity is not real or is due to an unidentified
nearby object.  In addition, the second brightest $90 \, {\rm \mu m}$ pixel
is the central one and the ratio of 60 to $90 \, {\rm \mu m}$ fluxes is
plausible for a real source located at the star.

The fact that the central pixel has the highest intensity for 11 out
of 12 stars at $60 \, {\rm \mu m}$ suggests that the far--IR IRAS sources
truly are associated with most of the giant stars presented
in Table 1 of \markcite{ZKL95}Zuckerman et al. (1995).

\begin{deluxetable}{lr@{$\, \pm \,$}lr@{$\, \pm \,$}lr@{$\, \pm \,$}lcr@{$\, \pm \,$}lr@{$\, \pm \,$}lr@{$\, \pm \,$}lcrrr}
\footnotesize
\tablecolumns{18}
\tablewidth{0pt}
\tablecaption{ISO and IRAS Fluxes (Jy)
\label{table:flux}}
\tablehead{
\colhead{} &
\multicolumn{6}{c}{$F_{\rm ISO}(60{\rm \mu m})$} &
\colhead{} &
\multicolumn{6}{c}{$F_{\rm ISO}(90{\rm \mu m})$} &
\colhead{} &
\multicolumn{3}{c}{} \\ \cline{2-7} \cline{9-14}
\colhead{} &
\multicolumn{2}{c}{center} &
\multicolumn{2}{c}{boundary} &
\multicolumn{2}{c}{background} &
\colhead{} &
\multicolumn{2}{c}{center} &
\multicolumn{2}{c}{boundary} &
\multicolumn{2}{c}{background} &
\colhead{} &
\multicolumn{3}{c}{$F_{\rm IRAS}$} \\ \cline{16-18}
\colhead{Star} &
\multicolumn{2}{c}{(pixel)} &
\multicolumn{2}{c}{(pixel sum)} &
\multicolumn{2}{c}{(average)} &
\colhead{} &
\multicolumn{2}{c}{(pixel)} &
\multicolumn{2}{c}{(pixel sum)} &
\multicolumn{2}{c}{(average)} &
\colhead{} &
\colhead{12${\rm \mu m}$} &
\colhead{25${\rm \mu m}$} &
\colhead{60${\rm \mu m}$}
}
\startdata
HD 119853	&	0.289 	&	0.089 	&	0.509 	&	0.185 	&	1.315 	&	0.059 	&&	0.187 	&	0.071 	&	0.229 	&	0.229 	&	1.047 	&	0.049 	&&	2.360 	&	0.571 	&	0.887 	\nl
HD 221776	&	0.271 	&	0.045 	&	0.293 	&	0.095 	&	0.630 	&	0.025 	&&	0.177 	&	0.061 	&	0.813 	&	0.112 	&	0.691 	&	0.049 	&&	5.080 	&	1.360 	&	0.840 	\nl
HD 153687	&	1.288 	&	0.127 	&	1.543 	&	0.414 	&	1.253 	&	0.069 	&&	0.755 	&	0.103 	&	0.669 	&	0.478 	&	1.349 	&	0.053 	&&	13.700 	&	4.160 	&	5.120 	\nl
HD 156115	&	0.265 	&	0.107 	&	0.225 	&	0.194 	&	1.648 	&	0.071 	&&	0.221 	&	0.103 	&	0.351 	&	0.191 	&	1.786 	&	0.060 	&&	8.900 	&	2.280 	&	0.913 	\nl
HD 202418	&	0.305 	&	0.036 	&	0.477 	&	0.135 	&	0.411 	&	0.019 	&&	0.218 	&	0.040 	&	0.444 	&	0.173 	&	0.558 	&	0.022 	&&	2.400 	&	0.696 	&	0.919 	\nl
HD 218559	&	0.157 	&	0.036 	&	0.320 	&	0.073 	&	0.406 	&	0.028 	&&	0.107 	&	0.030 	&	0.280 	&	0.066 	&	0.482 	&	0.018 	&&	3.570 	&	0.977 	&	0.580 	\nl
HD 212320	&	0.172 	&	0.083 	&	0.221 	&	0.080 	&	1.475 	&	0.045 	&&	0.180 	&	0.062 	&	0.436 	&	0.109 	&	1.007 	&	0.037 	&&	1.300 	&	0.422 	&	0.606 	\nl
HD 19745 	&	0.255 	&	0.034 	&	0.133 	&	0.083 	&	0.314 	&	0.025 	&&	0.220 	&	0.030 	&	-0.034 	&	0.054 	&	0.298 	&	0.020 	&&	0.296 	&	0.777 	&	0.612 	\nl
HD 24124 	&	0.000 	&	0.025 	&	0.145 	&	0.114 	&	0.351 	&	0.021 	&&	0.034 	&	0.021 	&	0.352 	&	0.293 	&	0.283 	&	0.016 	&&	0.334 	&	0.250 	&	0.400 	\nl
HD 111830	&	0.395 	&	0.044 	&	0.127 	&	0.100 	&	0.505 	&	0.022 	&&	0.337 	&	0.058 	&	0.398 	&	0.170 	&	0.776 	&	0.034 	&&	0.571 	&	0.231 	&	0.972 	\nl
HD 92253 	&	0.158 	&	0.032 	&	0.137 	&	0.062 	&	0.439 	&	0.019 	&&	0.139 	&	0.049 	&	0.198 	&	0.117 	&	0.770 	&	0.031 	&&	0.774 	&	0.223 	&	0.604 	\nl
HD 32440 	&	0.237 	&	0.033 	&	0.155 	&	0.095 	&	0.356 	&	0.021 	&&	0.154 	&	0.027 	&	0.073 	&	0.076 	&	0.444 	&	0.011 	&&	8.000 	&	2.000 	&	0.573 	\nl
\enddata
\tablecomments{
Fluxes are non--color corrected.  Central--pixel and sum of boundary--pixel
fluxes are listed in the second and third, and fifth and sixth column
for the 60 and 90~$\mu$m images, respectively.  The background intensity,
in an average background pixel, that was subtracted from the $\sf on$--source
fluxes is given in columns four and seven.  The IRAS 12, 25 and 60~$\mu$m
source fluxes are from the Faint Source Catalog.  The boundary pixel to the
south in the HD~221776 $90 \, {\rm \mu m}$ source image, which has the
out--of--center peak intensity, was not considered when deriving its
boundary--pixel flux.
}
\end{deluxetable}

\subsection{The Source Size}
\label{sec:size}

A $3 \times 3$ array is too coarse to give information on the source
size directly from its image.  However, if the distribution of the
emission source is simple and axisymmetric about the image center,
the extent of the source may be inferred from the ratio ($r$)
of the flux in the central pixel to the sum of the
fluxes in the 8 boundary pixels (Table \ref{table:flux}).
When defining $r$, only excess (i.e., non--photospheric) emission is considered.
By comparing the measured flux ratio $r_{obs}$ and a model
flux ratio $r_{mod}$, one may estimate the size of the extended source.

When calculating $r_{mod}$, one first needs a footprint, which is the
fraction of the energy in the PSF that falls onto a pixel as a function
of the pixel's location relative to the center of the PSF.
Then the fraction of the
flux onto a certain pixel from an extended emission source is obtained by
convolving the footprint with an assumed source distribution.  The
PIA package includes model footprints of the ISOPHOT C100 camera,
but a recent footprint calibration (\markcite{L99}Laureijs 1999) shows that
the observed footprints for the whole array at a few different locations
are slightly smaller than the PIA model values.  While the PIA model
gives the full 2--dimensional footprint, the calibration by Laureijs gives
footprint values only at few locations in the focal plane.  Thus one has
to modify the PIA model footprint so that it best matches the Laureijs
calibration.  We find that shrinking the x--y scale
(focal plane scale) of the model PIA footprint by 15~\% can well fit
the observed footprint values.  We choose altering the x--y scale of the
model footprint instead of dividing all footprint values
by a certain constant because the former also gives good fits to
the ratios of observed footprints while the latter does not.

\begin{figure*}
\centerline{\epsfxsize=16cm\epsfbox{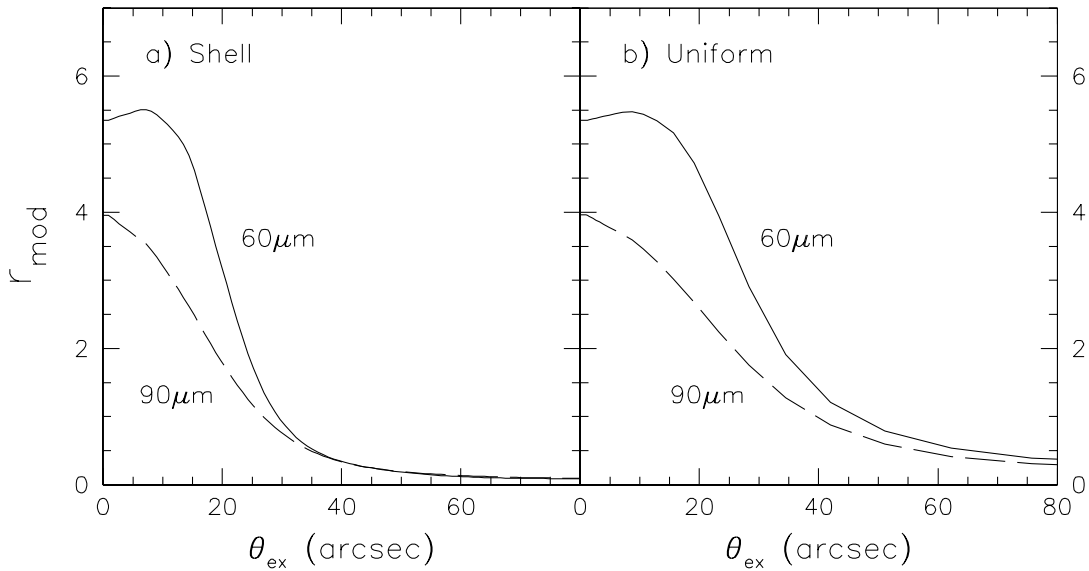}}
\caption
{\label{fig:psf}Modelled intensity ratio of central to boundary pixels
of the ISO C100 camera, $r_{mod}$, for the extended, excess far--IR emission
source at 60 ({\it solid lines}) and $90 \, \mu$m ({\it dashed lines}).
{\it a}) For an infinitesimally thin shell of circumstellar matter with
an angular radius of $\theta_{ex}$.
{\it b}) For a uniform distribution of cirrus material with
an angular radius of $\theta_{ex}$.  In case of the latter, the intensity
ratio is a function of central star's parameters, $D_*$, $T_*$, and $R_*$.
The intensity ratio shown in {\it b}) is for HD~119853.
$r_{mod}$ has an initial rise near $\theta_{ex}=0\arcsec$ at $60 \, \mu$m,
whose existence is a result of the detailed shape of the footprint.
}
\end{figure*}

Here we assume two simple models for the distribution of the material
responsible for excess far--IR emission: 1) an infinitesimally thin shell around
the central star with an angular radius of $\theta_{ex}$, appropriate for the
orbiting and ejected circumstellar dust models, and
2) uniform distribution of material centered on the star with an
angular radius of $\theta_{ex}$, appropriate for the cirrus hotspot model (see
\S~\ref{sec:introduction} and \markcite{J99}Jura 1999 for a detailed
description of the orbiting
dust, ejected dust, and cirrus hotspot models).  Then $r_{mod}$ is a ratio
of the footprint convolved with the assumed distribution
with a given $\theta_{ex}$ for central pixel ($f_c$) to
that for the sum of the boundary pixels ($f_b$):
\begin{equation}
\label{eq:rmod}
	r_{mod}(\theta_{ex}) \equiv {f_c (\theta_{ex}) \over f_b (\theta_{ex})}.
\end{equation}
For the shell distribution model, the footprint is convolved with
the projected distribution of a shell of constant intensity.  In the case
of a uniform distribution, the footprint is convolved with the
projected distribution of a filled dust sphere with a radial intensity
profile appropriate for optically thin, equilibrium
approximation\footnote{See Appendix B
of \markcite{Se85}Sopka et al. 1985 for the radial intensity profile.
The intensity profile is basically a Planck function with a local
temperature given by their equation~[B3], which is proportional to
$R^{-2/(4+p)}$ where $R$ is the radius from the star and $p$ is the
emissivity index.  We adopt $p=1.5$ following \markcite{J99}Jura 1999
and assume that the photospheric emission from the central star follows
a Planck function as well.}.
The calculated $r_{mod}(\theta_{ex})$ is presented in Figure~\ref{fig:psf}.

The observed fluxes also include a contribution from
the photosphere of the central star.
Following \markcite{J99}Jura (1999),
we estimate the photospheric emission from the star at $60 \, {\rm \mu m}$
using the IRAS $12 \, {\rm \mu m}$ flux and an assumed photospheric
ratio $F^* (60)/F^* (12)$ of 0.0371 (Jura obtained this value
by averaging the IRAS colors of the 9 brightest K and G giants in the
Yale Bright Star Catalog).
When converting the estimated IRAS photospheric flux for the ISO filters,
we adopt color--correction factors for a 4000~K blackbody.
The 90~$\mu$m photospheric flux is extrapolated from the 60~$\mu$m photosphere
with a blackbody assumption.  The photosphere emission at 60 and $90 \, \mu$m
is assumed to be not diluted by the presence of a circumstellar shell.
The estimated photospheric fluxes are given in Table~\ref{table:fluxest}.
The contribution of the central star to the flux in each pixel can
then be determined from the footprint and we subtract these photospheric
fluxes from the observed fluxes to determine $r_{obs}$:
\begin{equation}
\label{eq:robs}
	r_{obs} \equiv {F_c - P_c F^* \over F_b - P_b F^*},
\end{equation}
where $F$ is the observed source flux, $F^*$ is the estimated photospheric
flux, $P$ is the fraction of the flux that falls onto the central pixel or
boundary pixels from a point source located at the center of the array,
and the subscripts $c$ and $b$ are for the center and boundary pixels,
respectively.  $P_c$ and $P_b$ values, adopted from the modified
PIA footprint model, are 0.66 \& 0.12 for $60 \, {\rm \mu m}$,
and 0.59 \& 0.15 for $90 \, {\rm \mu m}$, respectively (the rest of the
flux falls outside the array).  The variables for the boundary pixels
are summed over the eight pixels.

The angular size of the extended emission source, $\theta_{ex}$,
can be obtained from equating $r_{mod}$ and $r_{obs}$.
We find $\theta_{ex}$ that satisfy $r_{mod}(\theta_{ex}) = r_{obs}$ for both
a shell and a uniform source distribution.
The size of the extended emission source, $R_{ex}$, then follows from
this $\theta_{ex}$ and the Hipparcos measured distance to the star,
$D_*$ (see Table~\ref{table:jura}).  Derived $R_{ex}$ values and
their uncertainties for our targets are given in Table~\ref{table:size}
as well as $r_{obs}$.

\begin{deluxetable}{lrrr}
\footnotesize
\tablecolumns{4}
\tablewidth{7cm}
\tablecaption{Color--Corrected Photospheric Fluxes (Jy)
\label{table:fluxest}}
\tablehead{
\colhead{} &
\multicolumn{3}{c}{$F^{*,cc}$} \\ \cline{2-4}
\colhead{Star} &
\colhead{(25)} &
\colhead{(60)} &
\colhead{(90)}
}
\startdata
HD 119853	&	0.393 	&	0.067 	&	0.030 	\nl
HD 221776	&	0.845 	&	0.144 	&	0.065 	\nl
HD 153687	&	2.280 	&	0.388 	&	0.174 	\nl
HD 156115	&	1.481 	&	0.252 	&	0.114 	\nl
HD 202418	&	0.399 	&	0.068 	&	0.031 	\nl
HD 218559	&	0.594 	&	0.101 	&	0.045 	\nl
HD 212320	&	0.216 	&	0.037 	&	0.016 	\nl
HD  19745	&	0.049 	&	0.008 	&	0.004 	\nl
HD  24124	&	0.056 	&	0.009 	&	0.004 	\nl
HD 111830	&	0.095 	&	0.016 	&	0.007 	\nl
HD  92253	&	0.129 	&	0.022 	&	0.010 	\nl
HD  32440	&	1.331 	&	0.227 	&	0.102 	\nl
\enddata
\end{deluxetable}

The $\theta_{ex}$ values found here for the shell distribution
(see Table~\ref{table:size}) range from $20\arcsec$ to $40\arcsec$
($30\arcsec$ to $70\arcsec$ for the uniform distribution),
implying that we were able to deconvolve extended sources whose
angular size is smaller than the $43\arcsec$ 1~pixel size of the detector.
\markcite{Pe97}Plets et al. (1997), whose
work was similar to that of \markcite{ZKL95}Zuckerman et al. (1995), state
that most of their luminosity class III giants with excess far
infrared emission appear to be unresolved in IRAS scans at
$60 \, {\rm \mu m}$.  We have convolved the $60 \, {\rm \mu m}$
IRAS PSF with the projected distribution of a shell with an angular radius
of $30\arcsec$, which is the typical angular size of extended emission sources
derived in the present study, and found that the full-width-half-maximum
of the convolved distribution was only $\sim 15$~\% larger than that of the
PSF.  This small difference is thought to be the reason that the sources in
Plets et al. ``look'' unresolved.  By deconvolving the PSF,
\markcite{HZ91}Hawkins \& Zuckerman (1991) were able to resolve
some objects with an angular radius smaller than
$30\arcsec$ at $60 \, {\rm \mu m}$, but their objects mostly had relatively
high fluxes (larger than a few Jy).  Since our targets have fluxes one to
two orders lower (mostly a fraction of $\sim 1$~Jy), deconvolving IRAS data
will not reliably resolve our targets (\markcite{H00}Hawkins 2000).  Thus
the ISO images appear to be the only ones currently available that can
resolve our targets in the far--IR.

\subsection{Robustness of Our Results}

We consider possibilities that could give a systematic bias to
the derived source sizes.  To estimate
the effects of inaccurate background determination, we decrease and
increase the background by 5~\%, which is the average background
uncertainty, and recalculate the $R_{ex}$ values.  While the decrease of
the background produces 10--20~\% increase in $R_{ex}$, the increase of the
background results in negative $F^{ex}$ values (at central or boundary pixel;
a negative $F^{ex}$ implies that the target flux is smaller than the
estimated photospheric flux).  Thus, while a systematic overestimation of
the background, if any, will result in slight underestimation of
$R_{ex}$, considerable underestimation of the background seems unlikely.
Moreover, as shown in Table~\ref{table:flux}
and Figure~\ref{fig:sky}, the uncertainty of the ISO background is only
few percent and there exists a good agreement between the ISO and IRAS
data without a bias toward a particular observation, which suggests that
a systematic over-- or underestimation in the ISO background determination
is very unlikely.  On the other hand, we find that our `second calibration'
(dividing the source flux by the background flux pixel--by--pixel) has an
effect of increasing $R_{ex}$ values only by $\sim 15$\%.

The estimation of $F^*$ also could be a possible cause of the bias,
if any.  Here we try two methods for estimating $F^*$ different
from the one used in \S~\ref{sec:analysis}.  The first method
(method A of \markcite{ZKL95}Zuckerman et al. 1995) estimates $F^*(12)$
using the empirical relation
between $F^*(12)$ and V magnitude as a function of B--V color by
\markcite{WCA87}Waters, Cot\'e, \& Aumann (1987) and extrapolates
to $F^*(60)$ and $F^*(90)$ for a blackbody spectrum.  The second
method (method B of \markcite{ZKL95}Zuckerman et al. 1995) assumes
the IRAS $12\,\mu$m flux is photospheric
(as in \S~\ref{sec:analysis}) and extrapolates to $F^*(60)$ and $F^*(90)$
for a blackbody spectrum.  The new $F^*$ estimation with methods A and
B gives a considerable change in $R_{ex}$ only for HD~156115 and HD~32440,
where $F^*$ is more than one--third of the total source flux.  These new
methods give $\sim 10$~\% larger $F^*$, and result in
10--20~\% $R_{ex}$ increase for the above two stars.  Thus the
method of estimating $F^*$ will not significantly affect the $R_{ex}$
values.

\begin{deluxetable}{lr@{$\, \pm \,$}lcr@{$\,$}lccr@{$\,$}lc
                    r@{$\, \pm \,$}lcr@{$\,$}lccr@{$\,$}l}
\footnotesize
\tablecolumns{20}
\tablewidth{0pt}
\tablecaption{Observed and Derived Size Parameters
\label{table:size}}
\tablehead{
\colhead{} &
\multicolumn{9}{c}{60~$\mu$m} &
\colhead{} &
\multicolumn{9}{c}{90~$\mu$m} \\ \cline{2-10} \cline{12-20}
\multicolumn{3}{c}{} &
\multicolumn{3}{c}{Shell} &
\colhead{} &
\multicolumn{3}{c}{Uniform} &
\colhead{} &
\multicolumn{2}{c}{} &
\multicolumn{3}{c}{Shell} &
\colhead{} &
\multicolumn{3}{c}{Uniform} \\ \cline{4-6}\cline{8-10}\cline{14-16}\cline{18-20}
\multicolumn{3}{c}{} &
\colhead{$\theta_{ex}$} &
\multicolumn{2}{c}{$R_{ex}$} &
\colhead{} &
\colhead{$\theta_{ex}$} &
\multicolumn{2}{c}{$R_{ex}$} &
\multicolumn{3}{c}{} &
\colhead{$\theta_{ex}$} &
\multicolumn{2}{c}{$R_{ex}$} &
\colhead{} &
\colhead{$\theta_{ex}$} &
\multicolumn{2}{c}{$R_{ex}$} \\
\colhead{Star} &
\multicolumn{2}{c}{$r_{obs}$} &
\colhead{($\arcsec$)} &
\multicolumn{2}{c}{(AU)} &
\colhead{} &
\colhead{($\arcsec$)} &
\multicolumn{2}{c}{(AU)} &
\colhead{} &
\multicolumn{2}{c}{$r_{obs}$} &
\colhead{($\arcsec$)} &
\multicolumn{2}{c}{(AU)} &
\colhead{} &
\colhead{($\arcsec$)} &
\multicolumn{2}{c}{(AU)}
}
\startdata
HD 119853	&	0.48 	&	0.23 	&	36 	&	4200	& $_{-	500	}^{+	1000	}$	&	&	66 	&	7700	& $_{-	1400	}^{+	14400	}$	&&	0.74 	&	0.80 	&	30 	&	3500	& $_{-	1000	}^{+	\infty	}$	&	&	46 	&	5300	& $_{-	1700	}^{+	\infty	}$	\nl
\tablevspace{1ex}
HD 221776	&	0.62 	&	0.23 	&	33 	&	7000	& $_{-	600	}^{+	1000	}$	&	&	57 	&	12000	& $_{-	1700	}^{+	3800	}$	&&	0.56 	&	0.32 	&	33 	&	7000	& $_{-	1100	}^{+	2500	}$	&	&	53 	&	11000	& $_{-	2200	}^{+	9100	}$	\nl
\tablevspace{1ex}
HD 153687	&	0.68 	&	0.19 	&	33 	&	4000	& $_{-	200	}^{+	400	}$	&	&	53 	&	6500	& $_{-	700	}^{+	1200	}$	&&	0.99 	&	0.72 	&	27 	&	3300	& $_{-	800	}^{+	2000	}$	&	&	39 	&	4800	& $_{-	1300	}^{+	5100	}$	\nl
\tablevspace{1ex}
HD 156115	&	0.46 	&	0.44 	&	36 	&	9300	& $_{-	1600	}^{+	\infty	}$	&	&	71 	&	18100	& $_{-	5500	}^{+	\infty	}$	&&	0.43 	&	0.31 	&	37 	&	9400	& $_{-	1700	}^{+	6900	}$	&	&	62 	&	15800	& $_{-	4000	}^{+	\infty	}$	\nl
\tablevspace{1ex}
HD 202418	&	0.55 	&	0.17 	&	35 	&	7100	& $_{-	500	}^{+	800	}$	&	&	63 	&	12900	& $_{-	1600	}^{+	4100	}$	&&	0.45 	&	0.19 	&	36 	&	7400	& $_{-	900	}^{+	1700	}$	&	&	60 	&	12300	& $_{-	2100	}^{+	6900	}$	\nl
\tablevspace{1ex}
HD 218559	&	0.28 	&	0.09 	&	43 	&	7700	& $_{-	700	}^{+	1200	}$	&	&	151 	&	27100	& $_{-	12200	}^{+	\infty	}$	&&	0.28 	&	0.10 	&	43 	&	7700	& $_{-	800	}^{+	1700	}$	&	&	84 	&	15100	& $_{-	3200	}^{+	\infty	}$	\nl
\tablevspace{1ex}
HD 212320	&	0.67 	&	0.41 	&	33 	&	4600	& $_{-	500	}^{+	1600	}$	&	&	56 	&	7900	& $_{-	1600	}^{+	19300	}$	&&	0.39 	&	0.17 	&	38 	&	5400	& $_{-	700	}^{+	1200	}$	&	&	65 	&	9200	& $_{-	1600	}^{+	7100	}$	\nl
\tablevspace{1ex}
HD  19745	&	1.89 	&	1.20 	&	24 	&	12200	& $_{-	2100	}^{+	4100	}$	&	&	39 	&	20000	& $_{-	5100	}^{+	21600	}$	&&	\multicolumn{2}{c}{\nodata}			&	\nodata	&	\multicolumn{2}{c}{\nodata}						&	&	\nodata	&	\multicolumn{2}{c}{\nodata}						\nl
\tablevspace{1ex}
HD 111830	&	3.07 	&	2.44 	&	20 	&	3200	& $_{-	3200	}^{+	2000	}$	&	&	28 	&	4500	& $_{-	4500	}^{+	5900	}$	&&	0.84 	&	0.39 	&	29 	&	4500	& $_{-	700	}^{+	1100	}$	&	&	45 	&	7000	& $_{-	3100	}^{+	3000	}$	\nl
\tablevspace{1ex}
HD  92253	&	1.06 	&	0.52 	&	29 	&	5600	& $_{-	600	}^{+	1100	}$	&	&	48 	&	9200	& $_{-	1600	}^{+	4700	}$	&&	0.68 	&	0.46 	&	31 	&	6000	& $_{-	1100	}^{+	3200	}$	&	&	50 	&	9600	& $_{-	2300	}^{+	46900	}$	\nl
\tablevspace{1ex}
HD  32440	&	0.63 	&	0.39 	&	33 	&	7300	& $_{-	900	}^{+	2700	}$	&	&	56 	&	12200	& $_{-	2400	}^{+	30900	}$	&&	1.51 	&	1.60 	&	22 	&	4800	& $_{-	2400	}^{+	\infty	}$	&	&	31 	&	6800	& $_{-	3600	}^{+	\infty	}$	\nl
\enddata
\tablecomments{Uncertainties are at 1--$\sigma$ level.
The uncertainty of $r_{obs}$ comes
from the source fluxes, $F_c$ and $F_b$, and the uncertainty of $R_{ex}$ 
corresponds to that of $\theta_{ex}$ which is propagated from $r_{obs}$.
We do not apply our size analysis to $90 \, \mu$m HD~19745 data
because they have negative $F_b$. HD~24124 is not listed here because
the far--IR emission near the star is thought to be due to a galaxy.}
\end{deluxetable}

\begin{deluxetable}{lcccccccrrrcrr}
\footnotesize
\tablecolumns{14}
\tablewidth{0pt}
\tablecaption{Model Parameters
\label{table:jura}}
\tablehead{
\multicolumn{9}{c}{} &
\multicolumn{2}{c}{$T_{ex,1}$} &
\colhead{} &
\multicolumn{2}{c}{$T_{ex,2}$} \\ \cline{10-11} \cline{13-14}
\colhead{} &
\colhead{Sp.} &
\colhead{${\rm m_V}$} &
\colhead{B--V} &
\colhead{$D_*$} &
\colhead{$|b|$} &
\colhead{$L$} &
\colhead{$T_*$} &
\colhead{$R_*$} &
\colhead{IRAS} &
\colhead{ISO} &
\colhead{} &
\colhead{IRAS} &
\colhead{ISO} \\
\colhead{Star} &
\colhead{Type} &
\colhead{(mag)} &
\colhead{(mag)} &
\colhead{(pc)} &
\colhead{($\arcdeg$)} &
\colhead{(${\rm L_\odot}$)} &
\colhead{(K)} &
\colhead{($10^{12}$cm)} &
\colhead{(K)} &
\colhead{(K)} &
\colhead{} &
\colhead{(K)} &
\colhead{(K)}
}
\startdata
HD 119853	&	G8 III	&	5.50 	&	0.90 	&	116	&	48	&	$8.69\times10^1$	&	5050	&	0.85 	&	54 	&	116 	&&	47 	&	68 	\nl
HD 221776	&	K7 III	&	6.18 	&	1.59 	&	208	&	22	&	$4.51\times10^2$	&	3800	&	3.43 	&	87 	&	90 	&&	70 	&	59 	\nl
HD 153687	&	K4 III	&	4.82 	&	1.48 	&	123	&	22	&	$3.86\times10^2$	&	4020	&	2.83 	&	81 	&	141 	&&	67 	&	74 	\nl
HD 156115	&	K5 III	&	6.52 	&	1.45 	&	255	&	13	&	$3.24\times10^2$	&	4070	&	2.53 	&	94 	&	48 	&&	75 	&	38 	\nl
HD 202418	&	K3 III	&	6.42 	&	1.41 	&	204	&	30	&	$2.10\times10^2$	&	4140	&	1.97 	&	77 	&	96 	&&	64 	&	61 	\nl
HD 218559	&	K4 III	&	6.43 	&	1.50 	&	179	&	35	&	$1.95\times10^2$	&	3980	&	2.05 	&	91 	&	76 	&&	73 	&	53 	\nl
HD 212320	&	G6 III	&	5.92 	&	1.00 	&	141	&	50	&	$9.53\times10^1$	&	4840	&	0.97 	&	82 	&	61 	&&	67 	&	46 	\nl
HD 19745 	&	K1 III	&	9.10 	&	1.05 	&	500	&	46	&	$6.67\times10^1$	&	4770	&	0.84 	&	148 	&	81 	&&	104 	&	55 	\nl
HD 24124 	&	K1 III	&	8.45 	&	1.31 	&	505	&	48	&	$1.69\times10^2$	&	4300	&	1.64 	&	100 	&	\nodata	&&	79 	&	\nodata	\nl
HD 111830	&	K0 III	&	7.78 	&	1.25 	&	156	&	15	&	$2.77\times10^1$	&	4400	&	0.63 	&	69 	&	82 	&&	58 	&	56 	\nl
HD 92253 	&	K0 III	&	7.42 	&	1.19 	&	192	&	15	&	$5.43\times10^1$	&	4500	&	0.85 	&	65 	&	75 	&&	55 	&	53 	\nl
HD 32440 	&	K4 III	&	5.47 	&	1.52 	&	218	&	34	&	$7.39\times10^2$	&	3950	&	4.06 	&	104 	&	66 	&&	81 	&	48 	\nl
\enddata
\tablecomments{This table is similar to Table 1 of \markcite{J99}Jura (1999).
$|b|$ is the absolute galactic latitude, $L$ is the bolometric luminosity,
and $T_{ex,1}$ and $T_{ex,2}$ are the grain temperatures for the orbiting
dust and and the ejected dust models, respectively.
The other parameters are defined in the text.  HD~153687 (30~Oph) and HD~212320
(HR~8530) are also included in Jura's list.  ${\rm m_V}$, B--V, and $D_*$ are
from the Hipparcos catalog except for HD 19745, for which the magnitudes in
the Tycho catalog were adopted and $D_*$ was determined photometrically with
${\rm M_V}=0.6$.  $T_{ex}$ from the ISO data of HD~24124 is not given because
its peak intensity is not on the central pixel.}
\end{deluxetable}

Inaccurate PSF or footprint could also affect our size analysis.
As mentioned earlier, we used the model footprint modified for a
recent recalibration based on real ISO data from the same C100 camera
(\markcite{L99}Laureijs 1999).
Recalculation of $R_{ex}$ with the original model footprint produces
a change in the results of less than 20~\% (the footprint modification
resulted in about 50~\% increase in $r_{mod}(\theta_{ex}=0)$, or $P_c$/$P_b$,
but the change in $r_{mod}$ near the $r_{obs}$ values of our targets
due to the modification was relatively small).
Thus unless the focal plane scale of the true footprint is
largely different from that of the adopted footprint, inaccuracy
in the adopted footprint is expected not to considerably affect
our derived $R_{ex}$.


\section{DISCUSSION}
\label{sec:discussion}


\markcite{J99}Jura (1999) estimated the extent of the circumstellar
material around 8 nearby giants with infrared excess using three
source models.  Since the models result in sizes that differ
from each other by almost an order of magnitude,
the size of the infrared excess source may be used to judge which model
is most consistent with the observations.

\begin{deluxetable}{lrrrr}
\footnotesize
\tablecolumns{5}
\tablewidth{0pt}
\tablecaption{Model $R_{ex}$ of Target Stars (AU)
\label{table:modelrex}}
\tablehead{
\multicolumn{3}{c}{} &
\colhead{Cirrus} &
\colhead{Cirrus} \\
\colhead{Star} &
\colhead{Orbiting} &
\colhead{Ejected} &
\colhead{(60)} &
\colhead{(90)}
}
\startdata
HD 119853	&	100 	&	2,100 	&	29,000 	&	88,000 	\nl
HD 221776	&	210 	&	3,000 	&	53,000 	&	160,000 	\nl
HD 153687	&	120 	&	2,300 	&	51,000 	&	160,000 	\nl
HD 156115	&	310 	&	4,200 	&	47,000 	&	140,000 	\nl
HD 202418	&	150 	&	2,400 	&	39,000 	&	120,000 	\nl
HD 218559	&	150 	&	2,200 	&	36,000 	&	110,000 	\nl
HD 212320	&	150 	&	2,200 	&	29,000 	&	89,000 	\nl
HD  19745	&	48 	&	780 	&	24,000 	&	74,000 	\nl
HD  24124	&	100 	&	1,200 	&	36,000 	&	110,000 	\nl
HD 111830	&	72 	&	1,100 	&	15,000 	&	45,000 	\nl
HD  92253	&	120 	&	1,800 	&	21,000 	&	64,000 	\nl
HD  32440	&	300 	&	4,000 	&	70,000 	&	210,000 	\nl
\enddata
\end{deluxetable}

Jura's models require the
temperature and radius of the target star ($T_*$, $R_*$) and the
temperature of the circumstellar grains ($T_{ex}$).  $T_*$ is
obtained from the (B--V) color and $R_*$ is inferred from $T_*$,
distance to the star $D_*$, and V magnitude (see Table~\ref{table:jura}).
\markcite{F96}Flower's (1996) conversion between (B--V) color and $T_*$,
and the assumption that $M_{bol}{\rm (Sun)} = 4.74$ (\markcite{BCP98}Bessel,
Catelli, \& Plez 1998) are used in the above calculation.

To estimate $T_{ex}$, excess far--IR fluxes at two passbands are fit by
$\nu B_\nu (T_{ex})$ for the ejected dust model, and by $B_\nu (T_{ex})$
for the orbiting dust model, where $B_\nu$ is the Planck function and
$\nu$ is the frequency (see \markcite{J99}Jura 1999).
Both IRAS and ISO data were used for the fitting,
but we fit the IRAS data ($25\,\mu m$, $60\,\mu m$) and the ISO data
($60\,\mu m$, $90\,\mu m$) separately because the two observations have
different beam sizes.  For the IRAS data, excess fluxes are calculated
with assumed photospheric ratios $F^* (25)/F^* (12) = 0.233$ and
$F^* (60)/F^* (12) = 0.0371$.  For the ISO data, we obtain the
excess fluxes by subtracting the photospheric flux calculated in
\S~\ref{sec:size} from the total flux.  The $T_{ex}$ values derived from
the two data sets are each shown in Table~\ref{table:jura}.
We use the average of the two for further analysis.

Jura's estimation for model source sizes are then given by
\begin{equation}
R_{ex} = \left \{
\begin{array}{l}
	 0.5 \, R_* \left ( \frac{T_*}{T_{ex}} \right )^2 \\
         0.5 \, R_* \left ( \frac{T_*}{T_{ex}} \right )^{2.5} \\
	 120 \, R_* \left ( \frac{k T_*}{h \nu} \right )^{2.75}
\end{array}
\begin{array}{l}
	 \, \, \, {\rm for \, orbiting \, dust;} \\
	 \, \, \, {\rm for \, ejected \, dust;} \\
	 \, \, \, {\rm for \, cirrus \, hotspot.}
\end{array}
\right .
\end{equation}
Note that $R_{ex}$ for the cirrus hotspot model is the radius at which
half the energy is emitted.
Table~\ref{table:modelrex} gives model $R_{ex}$ for our giants obtained
by the above equations.  Each model results in $R_{ex}$ that differ by
an order of magnitude, and only the cirrus model gives $R_{ex}$ dependent
on the wavelength.

Orbiting dust and ejected dust model $R_{ex}$ values are to be
compared to $R_{ex}$ values derived from observed $r_{obs}$ with an
assumption of the shell distribution for the excess emission source,
whereas the cirrus hotspot model $R_{ex}$ values are to be compared
to those derived with an assumption of the uniform distribution for
the excess emission source
(see Table~\ref{table:size} for the $R_{ex}$ derived from the
observations).

The region of 60 and 90 micron emission in the orbiting dust
(Kuiper--belt) model (Table~\ref{table:modelrex}) would be spatially
unresolved with the C100 detectors.  But Table~\ref{table:size} and
Figure~\ref{fig:psf} indicate that the 60 and 90~$\mu$m emission regions
of our target stars are resolved with typical angular radii $> 20 \arcsec$.
Thus, if the assumption of large (blackbody) grains in the
Kuiper--belt is appropriate, then the C100 data imply that the observed
far--IR emission is not due to orbiting dust.  If particles generated in a
Kuiper--belt structure are so small as to not radiate like blackbodies,
then the expected far--IR source size could be consistent with the observed
sizes.  However, as noted by \markcite{J99}Jura (1999), small particles
will be blown out of the systems by radiation pressure and would have
to be consistently replenished.  Then the dust masses required over
the lifetime of the phenomenon would be very large (see
\markcite{J99}Jura 1999).

Given the substantial uncertainties in the measured ratios for $R_{ex}$
(Table~\ref{table:size}), based on the ISOPHOT data, we are unable to choose
between the mass ejection and cirrus hotspot models (uncertainties
in $R_{ex}$ are asymmetrical and larger toward the positive direction).
The reason is that
the uncertainties of $r_{obs}$ are substantial, while the difference between
values of $r_{mod}$ calculated for these two models is not large.  Thus to
choose between them one must fall back on the types of arguments given by
\markcite{J99}Jura (1999).  Unfortunately, such arguments support neither
model particularly well.




\markcite{J99}Jura (1999) argued against the sporadic dust ejection
model for two reasons: (1) a recent ejection of matter would give
$F^{ex}(25)$ larger than $F^{ex}(60)$, but no K or late--G giants within 300~pc
of the Sun shows $F^{ex}(25) > F^{ex}(60)$; (2) one of the
giants with far--IR excess, $\delta$~And (HD~3627), is apparently expanding
at $v_{cs} \sim 300 \, {\rm km \, s^{-1}}$
(\markcite{JJR87}Judge, Jordan, \& Rowan-Robinson 1987),
implying only 20~yr for the dust to reach its estimated $R_{ex}$;
but none of the stars analyzed by \markcite{J99}Jura (1999) show significant
variability due to the expected dimming of starlight by the dust for the
first few months.
However, we find that 8 out of 92 giants in the list presented by
\markcite{ZKL95}Zuckerman et al. (1995)
have $F^{ex}(25) > F^{ex}(60)$, and such frequency agrees with
our simple calculation of $F^{ex}(25) - F^{ex}(60)$ evolution for the
detachment of a thin shell, composed of dust with $\nu B_\nu$ emissivity,
from the photosphere with a constant expanding velocity.
Furthermore, we note that $\delta$~And is an unusual K giant.
It is classified as a ``hybrid star'' which possesses both $10^{6-7}$~K
hot corona and cool stellar wind (see \markcite{HSR92}Haisch, Schmitt,
\& Rosso 1992 and references therein), and \markcite{JJR87}Judge et al.
(1987) suggested that the high velocity wind from $\delta$~And may not be
responsible for the formation of the circumstellar shell around the star.
If the dust shell is assumed to be blown away at $v_{cs}$ comparable to
that of the stellar wind of asymptotic giant branch stars (a few tens of
${\rm km \, s^{-1}}$), one would have a much lower possibility of finding
significant starlight variability in a given period.


Very recently, Kalas et al. (2000, in preparation) conducted coronagraphic
optical observations of 60 Vega-like stars (main-sequence stars with
apparent excess far-IR emission), and found reflection nebulae around
five stars which resemble those in the Pleiades.  This suggests the
cirrus hotspot model for the origin of excess far-IR emission from
those five stars.  Similar coronagraphic observations for our target
stars might help one in choosing between the mass ejection and cirrus
hotspot models.

\section{SUMMARY}
\label{sec:summary}

We have analyzed far infrared imaging data of 12 luminosity class III
stars with associated dust particles, observed with the C100 camera
on {\it ISO}.
Far--IR excess emission is associated with the central star for 11 targets,
and the excess emission of only one target appears to be due to a galaxy.
Thus we conclude that most of the stars presented by
\markcite{ZKL95}Zuckerman et al. (1995) actually heat
dust particles in their vicinity.

Three models for the origin of the circumstellar dust considered
by \markcite{J99}Jura (1999) predict very different source sizes.
To estimate the size of the far--IR emission source, we examined the
flux ratio of the central to eight boundary pixels of the $3\times3$
C100 array.
In one model considered by \markcite{J99}Jura (1999), the observed
dust is produced as ``Kuiper--belt'' like materials located within a few
hundred AU of the stars are warmed by the increasingly luminous giant
star (the orbiting dust model).  Such a far--IR emitting region would appear
spatially unresolved in the C100 images.
But the observed emission regions do appear to be spatially
resolved with radius at least a few thousand AU, and possibly significantly
larger.  With this size uncertainty, we are unable to choose between the
other two models discussed by Jura---sporadic mass ejection and interstellar
cirrus hotspot.  Neither of these models is in particularly good agreement
with all existing data (see, e.g., \markcite{J99}Jura 1999), so a clear
choice between them (or other models) awaits additional observations,
perhaps with SIRTF.



\acknowledgements
We are grateful to Carlos Gabriel and Rene Laureijs of the ESA--VILSPA,
Nanyao Lu of the IPAC, and Celeste Spangler for helping us with the ISO
data analysis.  We thank Eric Becklin and Mike Jura for helpful discussion
and reading the manuscript.
The ISOPHOT data presented in this paper was reduced using PIA,
which is a joint development by the ESA Astrophysics Division and
the ISOPHOT Consortium.  We have used the Simbad database, Aladin
sky atlas, and NED skyplot.




\end{document}